\documentclass[12pt]{article} 

\usepackage{subfig} 
\usepackage{epsfig} 
\usepackage{amsmath,amssymb}
\usepackage{graphicx}

\newcommand{\be}{\begin{equation}}
\newcommand{\ee}{\end{equation}} \newcommand{\bq}{\begin{eqnarray}} \newcommand{\eq}{\end{eqnarray}}
\newcommand{\f}{\frac}

\begin{document}

\title{Non-Cartesian Thinking} \author{Mikhail Kagan}  \vspace{.5cm} \maketitle

\begin{abstract} As we typically teach in an introductory mechanics course,
choosing a ``good'' reference frame with convenient axes may present a major simplification to a
problem. Additionally, knowing some conserved quantities provides an extremely powerful
problem-solving tool. While the former idea is typically discussed in the context of Newton's Laws,
the latter starts with introducing conservation of energy even later.

This work presents an elegant example of implementing both aforementioned ideas in the kinematical
context, thus providing a ``warm-up'' introduction to the standard tools used later on in dynamics.
Both the choice of the (non-orthogonal) reference frame and the conserved quantities are rather
non-standard, yet at the same time quite intuitive to the problem at hand. Two such problems are
discussed in detail with two alternative approaches. The first approach does not even require
knowledge of calculus. In the appendix, I also present the brute-force solution involving a coupled
system of differential equations. In addition, a few exercises and another similar problem for
students' ``homework'' are provided at the end.
\end{abstract}

\section*{Introduction} 
Chase problems were originally considered in the $17^{\rm th}$ century by Claude Perrault and later studied by Newton, Leibnitz, Huygens and others. In such a problem, there is typically a prey (leader) following a certain given trajectory and a pursuer whose trajectory ({\em curve of pursuit}) is to be found based on the specifics of the chasing algorithm. Chase problems have always received  attention of physics teachers as an opportunity to illustrate some concepts of kinematics. In general, the curve of pursuit follows from solving a system of differential equations describing the chase. However, there has been a number of papers (see e.g. \cite{Ship} and \cite{Dog}) that explain how, in some simple cases, the solutions can be  found by use of just basic calculus, without resorting to differential equations. In this paper, we shall see that some non-trivial questions can even be answered without calculus at all!

In fact, the goal of this paper is mostly pedagogical. It is not about solving a chase problem {\em per se}, but rather using the rabbit chase as a playground  for introducing two important ideas that go through the whole introductory physics sequence, as well as more advanced  courses: 
\begin{itemize} 
\item{\em Choosing a good reference frame can simplify a solution dramatically.}
From the very first course on (two- or higher-dimensional) kinematics, we are taught to use Cartesian coordinates as the
reference frame. Later on we get familiar with some other coordinates, such as polar (cylindrical) or spherical, which also belong to the class of {\em orthogonal} coordinate frames. While such orthogonal frames are certainly motivated by the problem at hand,  non-orthogonal frames hardly ever get mentioned. In this paper, a non-Cartesian frame arises rather naturally and leads to a major simplification.
\item{\em If there are conserved quantities, it may be a very powerful problem-solving tool.}

Speaking of physics in general, there are two technically different approaches to solving
problems: starting with equations of motion, compute the needed physical quantities {\bf or} avoid solving equations of motion explicitly by using conservations laws.  The latter method is based on that there are so-called {\em integrals of motion} - conserved quantities, like energy, momentum, angular momentum etc. - which can be used to relate
provided data to the unknowns of the given problem. 
\end{itemize}
Note that in a traditional introductory physics course, both ideas arise after $2^{\rm D}$-Kinematics. In the light of this, the problem explained in the paper may facilitate an early students' exposure to these fundamental concepts and enable them to think outside of the Cartesian box!

The paper is organized as follows. In the next two sections, two chase problems are solved using a non-trivial coordinate frame and non-trivial conserved  quantity. Note that the solutions do not require knowledge of calculus, so any student  familiar with concepts of uniform motion, vectors and basics of relative motion should be able to follow. After the chase problems I provide a few control questions about the solutions, as well as a seemingly unrelated ``homework'' problem that can be solved using the same approach. The ``brute-force'' solutions, based on differential equations, are provided in the Appendix.

\section*{Problem 1. Rabbit Chase.}
\begin{figure}[h!] 
\centerline{\includegraphics[clip=true, trim = 1.25in 18cm 5.4cm 3cm,width=6cm, keepaspectratio]{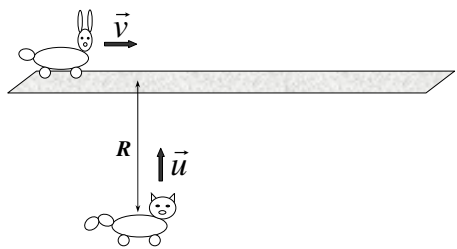}} \caption{Rabbit chase.\label{fig:Fox}} 
\end{figure}

{\em A rabbit is running along a straight road with constant speed  $v$. A fox is hiding  in the  bushes at distance $R$ from the road and starts chasing the rabbit 
when it is passing the point of nearest approach (see Fig \ref{fig:Fox}). The fox is always running at the rabbit with constant speed $u = v$.}\\

To anticipate students' question as to why the fox would try catching the rabbit in a such inefficient way, you do not really have to mention that the fox never took a basic kinematics course.
Foxes (as well as some other predators) really hunt like that, as they do not want to lose sight of their prey, especially in a forest environment. At this point, it could be pertinent to ask the students why this chase is inefficient, whether the fox will catch the rabbit and - once they say ``no'' - what will happen after a sufficiently long time. When they realize that in the ``long run'' the fox will end up on the road right behind the rabbit, running with the same speed hence at the same distance from its prey, you can ask the main question of the problem:\\

{\em After a long time, what will be the distance between the fox and the rabbit?}

As the fox's trajectory is highly non-trivial (see Eq.(\ref{Curve_of_Pursuit}) in the Appendix), you might want to discourage your students from trying to figure the shape of its path out and look for an elegant solution. 
The solution itself is as follows.
\begin{figure}[h] 
\centerline{\includegraphics[clip=true, trim = 1in 8in 4.25in 1in,width=8cm, keepaspectratio]{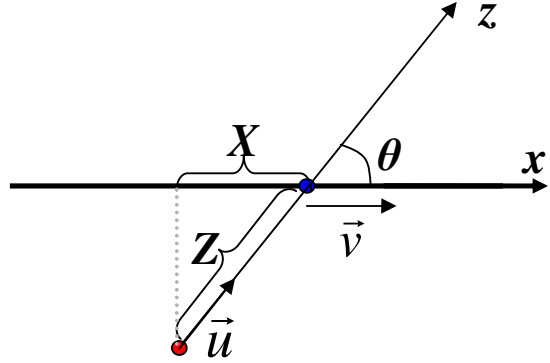}} \caption{The chase as viewed from the rabbit's reference frame. The rabbit is sitting at the origin, the $x$-axis is pointing along its path, and the $z$-axis is passing through the fox and the rabbit. The angle that the $z$-axis makes with respect to the $x$-axis, $\theta$, is changing with time from $\f{\pi}{2}$ down to 0. $Z$ is the distance between the fox and the rabbit, whereas $X$ is their horizontal separation.\label{fig:frame}} 
\end{figure}
First of all, it is convenient to consider the chase in the reference frame associated with the rabbit with the $x$-axis pointing along the the rabbit's path. The second axis, $z$, passes through the fox and the rabbit and makes a {\em variable} angle $theta$ with the $x$-axis (Fig. \ref{fig:frame}). In this frame, the rabbit is sitting at the origin, while the fox is following some trajectory that starts at $(0,-R)$ and ends somewhere on the $x$-axis. Denote the distance between the fox and the rabbit as $Z$ and their horizontal separation as $X$. 

Let us now consider how these two quantities change over a short time interval $\Delta t$. The rabbit is running to {\em increase} $X$ at the speed $v$, while the fox is trying to {\em decrease} the distance, but not as efficiently so. The fox's contribution to the decrease of $X$ is related to the horizontal component of its speed  $(u\cos\theta)$. Thus
\be\label{delta_X}
\Delta X = \left(v-u\cos\theta\right)\Delta t.
\ee
Similarly, during $\Delta t$ the distance between the fox and the rabbit will change by
\be\label{delta_Z}
\Delta Z = \left(-u+v\cos\theta\right)\Delta t.
\ee
Since $u=v$, the righthand sides of these equations are exact opposites of one another. Therefore, the sum $\Delta X + \Delta Z=0$, which implies that the quantity 
\[
Z+X=const.
\]
In the beginning, $X=0$ whereas $Z=R$, which fixes the constant to be $R$. After a long time, when the fox is right behind the rabbit, the two axes coincide, hence $X=Z$, which immediately yields
\be\label{FR_final_distance}
X_f=Z_f=R/2.
\ee
\section*{Problem 2. Missile vs Plane.}
No animals were hurt in the previous problem. Let us consider a similar  situation:
\\

\begin{figure} \centerline{\includegraphics[clip=true, trim = 2.5cm 18.2cm 5cm 2.5cm,width=8cm, keepaspectratio]{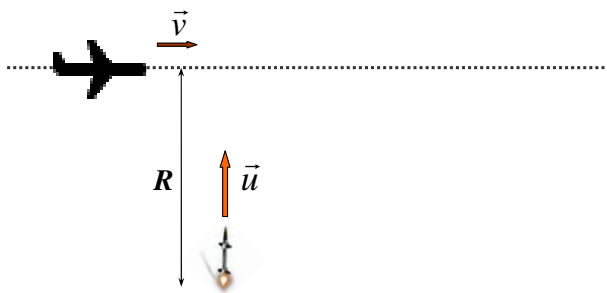}} \caption{Missile vs Plane.\label{fig:Missile}} 
\end{figure}

{\em A plane is following a straight horizontal trajectory at altitude $R$ with constant speed  $v$. A missile is launched from the ground when the plane is passing over it. The missile is always flying at the plane with constant speed $u$ ($u> v$) (see Fig \ref{fig:Missile}). How long will it take for the missile to hit the plane?}
\\

The problem can be solved using the same $(x,z)$-reference frame with the plane being at the origin. Although the speeds $v$ and $u$ are no longer equal. Nevertheless, 
Eqs. (\ref{delta_X}) and (\ref{delta_Z}) will still do the trick. Multiplying Eq.(\ref{delta_X}) by $v$ and adding Eq.(\ref{delta_Z}) multiplied by $u$ leads to 
\[
\Delta\left(vX+uZ\right)=\left(v^2-u^2\right)\Delta t.
\] 

Importantly, the coefficient in front of $\Delta t$ is a constant, so the quantity inside the parenthesis on the lefthand side is changing at a  constant rate. Therefore, the {\em overall} time interval can be  found by taking the {\em overall} change  $\left(v\Delta X+u\Delta Z\right)$ divided by $(v^2-u^2)$:
\be\label{Delta_t}
\Delta t=  \f{Ru}{u^2-v^2},
\ee
since $\Delta X= 0$ and $\Delta Z=-R$. At this point your students could verify that the answer has the right units. Moreover in the ``Fox \& Rabbit'' limit $u \rightarrow v$, the amount of time becomes infinite, so that the fox never catches the rabbit. Finally, if the missile is very much faster than the plane $u\gg v$, it is easy to see that the catch up time is  simply $\Delta t = R/u$, as if the plane was  not moving at all.

\section*{Discussion}
The solution did not require any knowledge of calculus. Students should know all the physical and mathematical ingredients of the method by the the time  they have covered vectors and relative motion. Having students work on such problems should help to facilitate the idea that a reference frame is a {\em tool} for describing and analyzing motion, rather than a fixed entity with the horizontal and vertical axes $x$ and $y$. And as a tool, it can be chosen differently and appropriately for each situation. We typically say that the reference frame should be {\em convenient}, which, with little to no experience, may sound a bit too abstract to students. In the case at hand, however, the choice is quite intuitive and appealing. Later on, when dealing with Newton's Laws and especially with objects on inclines (or swinging objects), it should be much more obvious for students that a {\em tilted} reference frame would be handy.

Another important idea that students should take home is the concept of conserved  quantities. While in this chase problem the constancy of $\left(vX+uZ\right)$ might look as a mere  coincidence and did not have a transparent physical meaning, as a rule conserved quantities do have a physical interpretation. Thus the fox and rabbit problem may become a natural introduction to the later conservation laws for energy, momentum and angular momentum.

There is, of course, a brute-force approach to the problem. That requires solving a coupled system of two differential equations and can be found in the appendix. Remarkably, the equations can be solved completely analytically. The solution also yields the trajectory equation for the fox/missile, which is a highly non-trivial curve. Interestingly, the knowledge of the trajectory is {\em not} necessary to answer the questions, which makes the elegant solution exist in the first place. It may also be curious to offer this problem not only to freshmen (pre-calculus students), but to more senior students who have taken an ODE course.  

It would not be a pedagogical paper, if not for the ``homework assignment''. In order to reinforce students' understanding of the solution explained above, you can offer them control questions on the following page. In addition to the control questions, you can challenge your students with a different problem that can be cracked using the same two ideas: a non-Cartesian reference frame and the concept of conserved quantities. 
\pagebreak
\section*{Homework}
\subsection*{Control Questions}
1) In the ``Fox \& Rabbit'' problem, how will the answer change if 
\begin{itemize}
\item the fox starts chasing the rabbit earlier/later: when the rabbit is at distance 	$R/2$ from the perpendicular?
\item when the fox takes off, the rabbit starts running at angle $\theta$ with respect to the  $x$-axis?  
\end{itemize}
2) In the Missile vs. Plane problem, how will the answer change if
\begin{itemize}
\item the missile is launched earlier/later: when the plane is at distance $R/2$ from the perpendicular?
\item when the missile is launched, the plane starts flying at angle $\theta$ with respect to the $x$-axis?
\end{itemize}

\subsection*{Challenge Problem}  
\begin{figure}[h]
\centerline{\includegraphics[clip=true, trim = 2.2cm 20.5cm 5.2cm 2cm,width=8cm, keepaspectratio]{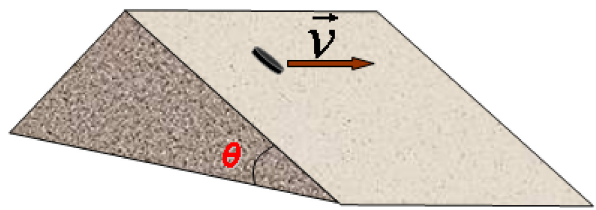}} \caption{A puck on an incline.\label{fig:puck}} 
\end{figure}

A hockey puck on a rough wedge with a coefficient of friction $\mu=\tan\theta$  receives a horizontal kick (see Fig \ref{fig:puck}). Assume that the wedge is very large and answer the questions below. 
\begin{itemize}
\item What will happen after a long time?
\item If the initial speed of the puck is $v$,  what is its final speed?
\end{itemize}
\pagebreak

\section*{Appendix} In this appendix, we consider the ``brute-force'' approach to the problem, which
amounts to deriving and then solving differential equations. As before, it is convenient to begin
consideration in the reference frame associated with the rabbit and introduce the $x$- and $z$-axes.

\begin{figure} 
\centerline{\includegraphics[clip=true, trim = 1in 8in 4.25in 1in, width=8cm,keepaspectratio]{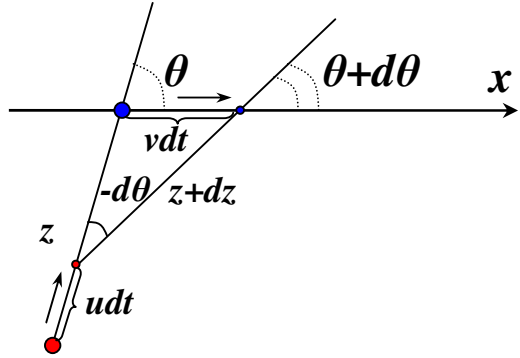}} 
\caption{Positions of the fox and rabbit at an arbitrary instant of time (big dots) and $dt$ later (small dots). During the $dt$, the fox's trajectory can be considered straight, so that the fox covers $udt$, while the rabbit moves  by $vdt$.The distance between the fox and the rabbit $z$ and the angle between the axes $\theta$ change by $dz$ and $d\theta$($<0$) respectively.\label{fig:dz}} 
\end{figure}

Fig \ref{fig:dz} depicts the chase at two infinitesimally close instants of time: $t$ and $t+dt$.
The corresponding separations between the fox and the rabbit are denoted $z(t)\equiv z$ and
$z(t+dt)\equiv z+dz$ respectively. Applying the law of sines to the triangle formed by $z-udt$,
$z+dz$ and the segment $vdt$ covered by the rabbit during $dt$, up to higher order terms we obtain

\[ z d\theta = - v\sin\theta dt, \] 
which results in the differential equation for $\theta$ 
\begin{equation}\label{Theta_dot} 
\dot\theta = -\frac{v \sin \theta}{z}. 
\end{equation} 
Rewriting Eq.(\ref{delta_Z}) as an explicit differential equation 

\be\label{Z_dot} \dot z = v\cos\theta - u, \ee we obtain a system of two ODE's for the two unknown
functions $z(t)$ and $\theta(t)$. The trajectory equation, $z(\theta)$, can be derived by dividing
Eq.(\ref{Z_dot}) by Eq.(\ref{Theta_dot}), which results in

\be\label{dz_dtheta} 
\frac{dz}{d\theta} = -z\frac{v\cos\theta - u}{v\sin\theta}\equiv
\frac{z}{\sin\theta}\left(1-\cos\theta+\delta\right), 
\ee 
where $\delta\equiv\frac{u}{v}-1$. Using the identity

\[ \frac{1-\cos\theta}{\sin\theta}\equiv\frac{\sin\theta}{1+\cos\theta}, \] 
Eq.(\ref{dz_dtheta}) can be easily integrated by separation of variables: 
\[ \ln z = -\ln |1+\cos\theta| + \delta
\ln\left|\tan\frac{\theta}{2}\right|+const. \] 
The constant of integration is fixed by the initial conditions. Specifically at $t=0$, $z=R$ and $\theta=\pi/2$. Thus we arrive at the fox's trajectory equation in the rabbit's reference frame 

\be\label{z_theta}
\frac{z}{R}=\frac{\left(\tan\frac{\theta}{2}\right)^\delta}{1+\cos\theta}. 
\ee 

Notice that the
parameter $\delta$ shows by how much the chaser is faster than the prey. For instance, $\delta=0$
corresponds to $u=v$, i.e. the ``Fox \& Rabbit'' problem. The ``Missile vs Plane'' situation is
characterized by a $\delta>1$.

\begin{figure}
\centerline{\includegraphics[clip=true, trim = 2cm 12cm 5cm 2cm, width=8cm, keepaspectratio]{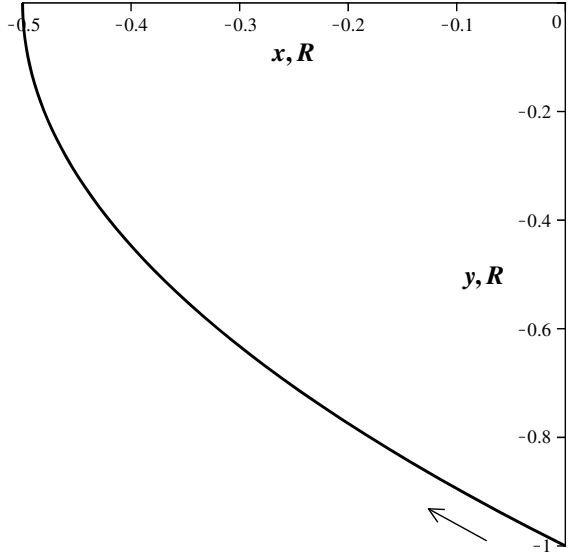}} 
\caption{In the reference frame associated with the rabbit, the fox is running along a parabolic trajectory from $(0, -R)$ to the top of the parabola at $(-R/2, 0)$. The rabbit is sitting at the focus of the parabola $(0, 0)$.}\label{fig:FR_x_vs_y}
\end{figure}

It is now easy to see what kind of a trajectory the fox is following. Setting $\delta=0$ in
Eq.(\ref{z_theta}) yields \[ z=\f{R}{1+\cos\theta}, \] which is nothing else but a parabola
equation written in polar coordinates. In the rabbit's reference frame, the rabbit sits in the focus
of the parabola at $(0, 0)$, while the fox starts off at $(0, -R)$ and runs toward the top of the parabola at
$(-R/2, 0)$, as shown in Fig.\ref{fig:FR_x_vs_y}. Hence the final distance between the fox and the rabbit is indeed $R/2$, as was found earlier.

\begin{figure} 
\centerline{\includegraphics[clip=true, trim = 2cm 12cm 5cm 2cm, width=8cm, keepaspectratio]{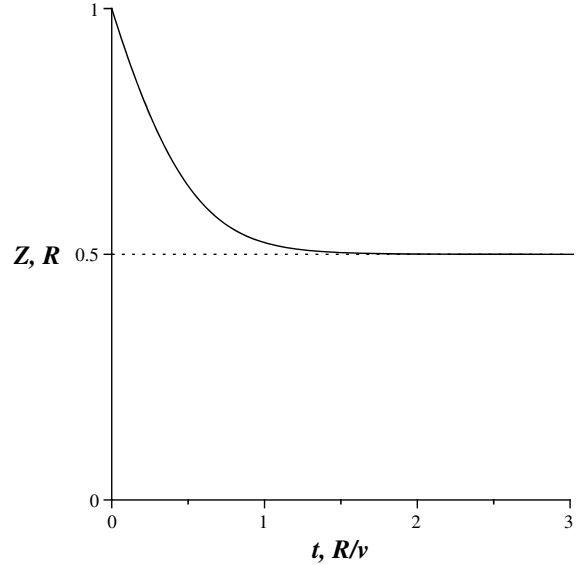}} 
\caption{Time dependence of the separation between the fox and the rabbit. The distance approaches its asymptotic value, $R/2$, exponentially fast.}
\label{fig:FR_z_vs_t} 
\end{figure}

The chase, however, is an asymptotic process, and it takes an infinite amount of time for the fox to get to the top. It is therefore insightful to consider the time dependence of the relevant coordinates. Substituting $z(\theta)$ from Eq.(\ref{z_theta}) back into Eq.(\ref{Theta_dot}) and separating the variables ($t$ and $\theta$) in the resulting equation leads to

\be\label{t_beta} 
t=\f{R}{2v}\left(\f{1-\beta^\delta}{\delta}+\f{1-\beta^{\delta+2}}{\delta+2}\right),
\ee
where $\beta=\tan(\theta/2)$. During the chase, $\theta$ goes from $\f{\pi}{2}$ down to $0$, which corresponds to $\beta$ ranging from $1$ to $0$ respectively. When arriving at the equation above, the constant of integration was fixed by requiring $t=0$ at $\beta=1$. Note that in terms of $\beta$ the separation between the fox and the rabbit (\ref{z_theta}) can be conveniently rewritten as
\be\label{z_beta}
z=\f{R}{2}\beta^\delta\left(1+\beta^2\right)
\ee
The last two equations thus describe the (parametric) time dependence of the fox-rabbit distance. Restricting Eq.(\ref{t_beta})  to $\delta=0$ amounts to applying L'Hopital's rule to the first term, resulting in
\[
t=\f{R}{2v}\left(-\ln\beta+\frac{1}{2}\left(1-\beta^2\right)\right),
\] 
in addition to
\[
z=\f{R}{2}\left(1+\beta^2\right).
\]
 It easy to see that after a long time ($\beta\rightarrow0$), the second term in the $t$-equation is negligible compared to the first one. This implies that $z$ approaches its  asymptotic value exponentially fast:
\[
\left(z-\f{R}{2}\right)\propto \exp{\left(-\f{4vt}{R}\right)}.
\]
The complete time  dependence of the fox-rabbit distance is displayed in  Fig.\ref{fig:FR_z_vs_t}.

\begin{figure} 
\centering 
\subfloat[Part 2][Time dependence of the missile-plane distance for (right to left) \\$\delta=0.1, 0.2, 0.5, 1.0, 2.0,$  and $3.0$. ]
{\includegraphics[clip=true, trim = 2.5cm 10.5cm 2cm 2cm, width=6.5cm, keepaspectratio]{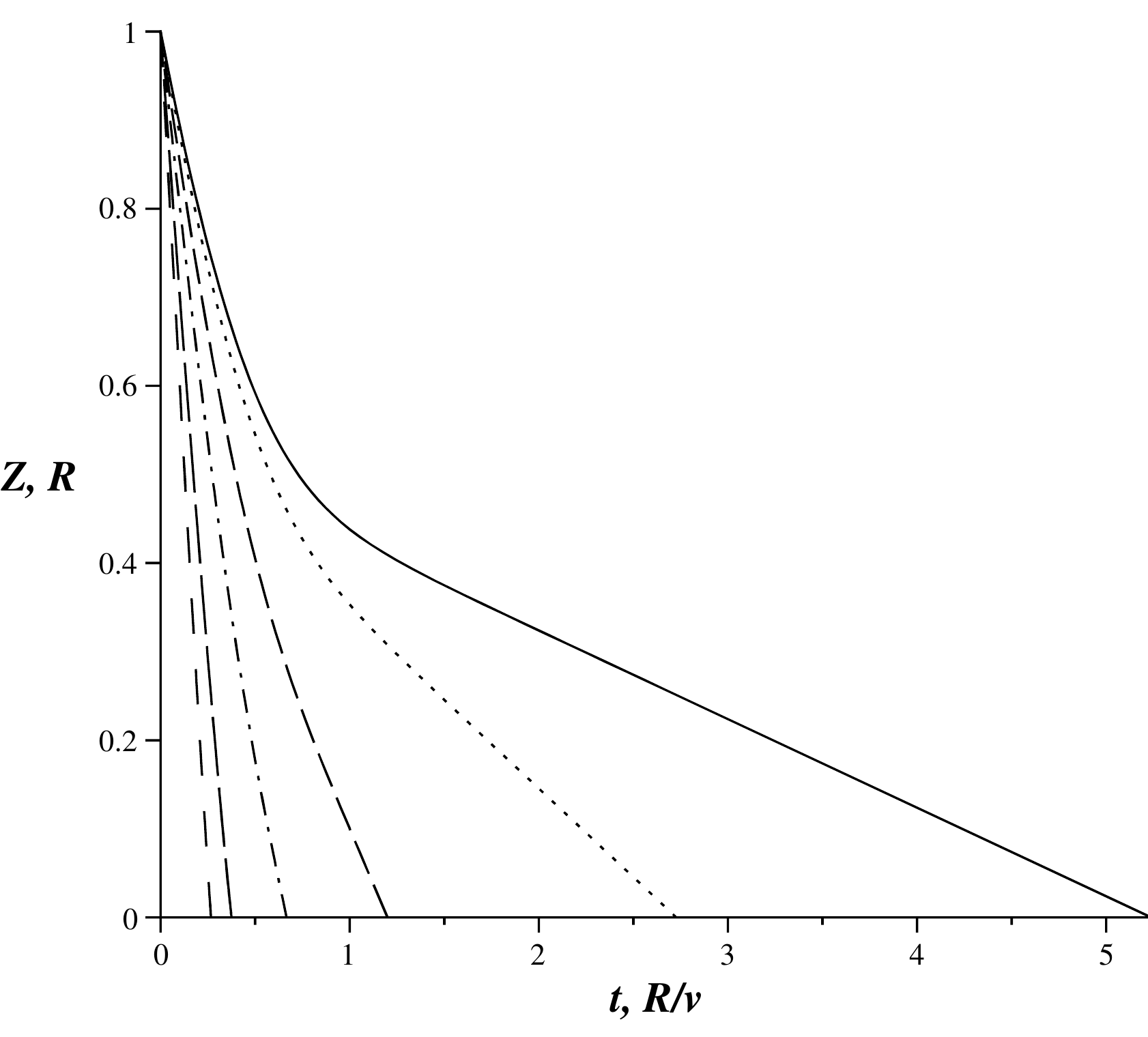}\label{fig:MP_z_vs_t}}\hfill
\subfloat[some caption][Missile's trajectory in the plane's reference frame for (left to right) \\$\delta=0.1, 0.2, 0.5, 1.0, 2.0,$  and $3.0$.]
{\includegraphics[clip=true, trim = 2.5cm 10.3cm 2cm 1cm, width=6.5cm, keepaspectratio]{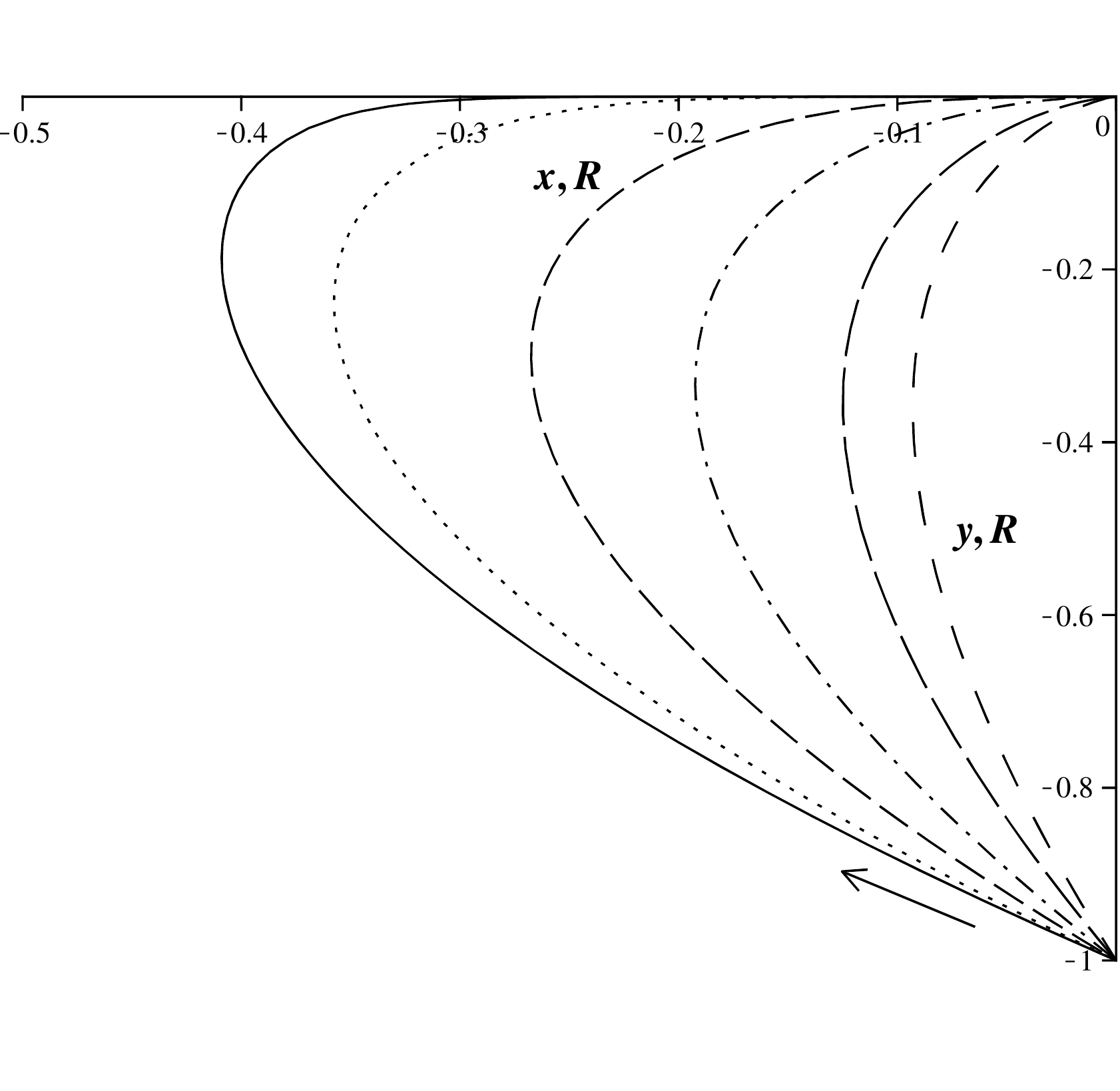}\label{fig:MP_x_vs_y}} 
\caption{Missile vs Plane graphs.}
\label{fig:MP} \end{figure}

We can repeat the same analysis for the ``Missile vs Plane'' problem. A set of parametric plots $z(t)$ for several non-zero values of $\delta$ is presented in Fig \ref{fig:MP_z_vs_t}. The rightmost curve ($\delta=0.1$) corresponds to a missile flying barely faster than the plane, i.e. $u=1.1v$. The leftmost curve ($\delta=3$) describes a very fast missile, whose speed is four times that of the plane. Clearly, all these curves reach $z=0$ in a finite amount of time that follows from Eq. (\ref{t_beta}). Setting $\beta=0$, we obtain
\be\label{catch_time}
t=\f{R}{v}\cdot\f{\delta+1}{\delta(\delta+2)}\equiv \f{Ru}{u^2-v^2},
\ee 
which agrees with the earlier result (\ref{Delta_t}). 

Possible trajectories of the missile are also of interest. As before, it is most conveniently done in the reference frame associated with the plane. Importantly all such trajectories (for $\delta>0$) terminate at the origin, as shown in Fig \ref{fig:MP_x_vs_y}. The leftmost curve corresponds to the slowest missile ($\delta=0.1$), whereas the rightmost curve describes the fastest case ($\delta=3$). Remarkably, the trajectory equation (\ref{z_beta}) allows for an explicit expression of the chaser's $(x, y)$-coordinates one in term of the other. In the prey's reference frame
\[
x=-z\cos\theta, \quad y=-z\sin\theta.
\]
By virtue of the standard trigonometric identities
\[
\cos\theta = \f{1-\beta^2}{1+\beta^2} \quad {\rm and} \quad \sin\theta =\f{2\beta}{1+\beta^2},
\]
the coordinates take the form
\be\label{x_and_y}
x=-\f{R}{2}\beta^\delta\left(1-\beta^2\right), \quad y=-R\beta^{\delta+1}.
\ee
Expressing $\beta$ in terms of $y$ from the second equation and substituting it into the first equation yields the trajectory equation
\bq
\left|\f{x}{R}\right|&=&\f{1}{2}\left(\left|\f{y}{R}\right|^{\f{\delta}{\delta+1}}-\left|\f{y}{R}\right|^{\f{\delta+2}{\delta+1}}\right), \nonumber\\
&\equiv& \f{1}{2}\left(\left|\f{y}{R}\right|^{1-\f{v}{u}}-\left|\f{y}{R}\right|^{1+\f{v}{u}}\right) \quad {\rm for} \quad x,y<0.\label{x_vs_y}
\eq
In particular, setting $\delta=0$ ($u=v$) recovers the parabolic path of the ``Fox \& Rabbit'' problem. If you are looking for a way of discouraging your students to follow the brute force solution and Eq.(\ref{x_vs_y}) does not look scary enough, I would suggest to present the chaser's trajectory equation in the laboratory reference frame ({\em curve of pursuit}) $x_{\rm Lab}(y)$, with $x_{\rm Lab}=x+vt$, where the time variable can be expressed in terms of $y$ using Eqs.(\ref{t_beta}) and (\ref{x_and_y}). The curve-of-pursuit equation reads
\be\label{Curve_of_Pursuit}
\f{x_{\rm Lab}}{R}=\f{uv}{u^2-v^2}+ \f{1}{2}\left(\f{u}{u+v}\left|\f{y}{R}\right|^{1+\f{v}{u}}-\f{u}{u-v}\left|\f{y}{R}\right|^{1-\f{v}{u}}\right) \quad {\rm for} \quad y<0.
\ee

\end{document}